\documentclass[twocolumn, prb]{revtex4}
\usepackage{graphicx}
\usepackage{latexsym}

\newcommand{\be}{\begin{equation}}
\newcommand{\ee}{\end{equation}}
\newcommand{\bea}{\begin{eqnarray}}
\newcommand{\eea}{\end{eqnarray}}

\begin{document}

\title{Conductivity of a suspension of nanowires in
a weakly conducting medium}
\author{Tao Hu, A. Yu. Grosberg, B. I. Shklovskii}
\affiliation{Department of Physics, University of Minnesota \\
116 Church Street SE, Minneapolis, MN 55455}
\date{\today}

\begin{abstract}
We study macroscopic electrical conductivity of a composite made
of straight or coiled nanowires suspended in poorly conducting
medium. We assume that volume fraction of the wires is so large
that spaces occupied by them overlap, but there is still enough
room to distribute wires isotropically. We found a wealth of
scaling regimes at different ratios of conductivities of the wire,
$\sigma_1$ and of the medium, $\sigma_2$, length of wires, their
persistent length and volume fraction. There are large ranges of
parameters where macroscopic conductivity is proportional to
$(\sigma_{1}\sigma_{2})^{1/2}$. These results are directly
applicable to the calculation of the macroscopic diffusion
constant of the nonspecific DNA-binding proteins in semi-dilute
DNA solution.
\end{abstract}
\maketitle

\section{Introduction}

Equilibrium and transport properties of composites are of great
interest because of their importance both in nature and technology.
One usually wants to characterize the composite macroscopically,
determining its effective properties such as conductivity,
dielectric constant, magnetic permeability etc., in terms of
properties of the respective constituents. Most of the theoretical
literature on this subject dealt with spherical (or single scale)
inclusions
\cite{Dykhne,Efros,Bergman,Niklasson,Clerc,Stauffer,Shin}. At the
same time, Monte-Carlo simulations and experiments
reveal\cite{Balberg,Balberg2,Balberg3,Foygel,Zuev,Kolesnikov,Yagil}
that elongated, needle or stick like inclusions, can be very
effective in modifying the properties of materials even at small
volume fractions. For example, composites made of metallic wires
with aspect ratio $a/l \ll 1$ ($a$ is the radius of the wire and $l$
is its length) immersed in a good insulator exhibit record values of
the dielectric constant \cite{Sarychev}. The transport properties of
such composites were studied in many works
\cite{Balberg,Foygel,Zuev,Kolesnikov}. A comprehensive review of
these works and a thorough study of the dielectric response of
conducting stick composites can be found in Ref. \cite{Sarychev},
but only in the asymptotic regime of very large conductivity of
wires.

Recently, another system attracted a lot of attention. It consists
of carbon nanotubes dispersed in ceramic or plastic material. It
was shown that nanotubes can greatly enhance the electrical and
thermal conductivities of the material
\cite{Andrews,Bryning,Choi,Park,Kilbride,Foygel}.

In this paper we present scaling theory of the macroscopic
conductivity $\sigma$ of the suspension of well conducting wires
with conductivity $\sigma_1$ in poorly conducting medium with a
finite conductivity $\sigma_2 \ll \sigma_1$. The wires can be
rigid sticks or flexible and coiled like conducting polymers. We
imagine that they are dispersed, randomly oriented and frozen in
the medium.

On the first glance, the problem of macroscopic conductivity of
the composite seems to belong to the realm of percolation
\cite{Stauffer}. Indeed, this would be true for the conducting
wires in perfectly insulating medium, $\sigma_2=0$, where
macroscopic conductivity could only be realized through the direct
contacts between wires.  In this article, we are interested in a
different problem - we assume that the medium does conduct,
$\sigma_2 \neq 0$, albeit poorly ($\sigma_2 \ll \sigma_1$). In
this case, although overall macroscopic current is carried mostly
along the wires, it is still able to switch from wire to wire
through the medium, depending on the random disordered
configuration of the wires. Accordingly, we mostly consider volume
fraction of wires $\phi$ to be not only small $\phi \ll 1$, but
actually smaller than the corresponding percolation threshold
$\phi \sim a/l$, such that the direct contacts between wires are
rare and completely negligible. We show later how our results
properly cross over to that of percolation at larger $\phi$.

For very dilute system of wires, when the distance between wires
is much larger than the length of the wire, the effective
conductivity $\sigma$ is trivially close to the conductivity of
the medium $\sigma_2$.  We therefore mostly deal with larger
concentrations with $\phi > a^2/l^2$ where the spheres containing
each wire strongly overlap (see Fig. \ref{fig:network}). In the
parlance of polymer science, we study the \emph{semidilute} system
\cite{Redbook} of wires. In terms of increasing concentration, our
theory continues as long as there remains enough room to
distribute wires isotropically. We show that although $\phi$ is
small in semi-dilute system of wires, macroscopic conductivity
$\sigma$ is dramatically enhanced when compared to $\sigma_2$.

The useful image to think about is a single typical current line
in the system.  It follows inside one wire for a long distance and
then bridges to a neighboring wire over a more-or-less narrow gap
in the medium, and then continues again in the wire for a long
distance. We denote $\lambda$ the length scale of one continuous
stretch of current line in one wire, it can be called
\emph{correlation} length. This is the key concept of the paper.
With the increase of wire conductivity $\sigma_1$, the correlation
length $\lambda$ increases, and so does macroscopic conductivity
$\sigma$, until $\lambda$ gets as large as the wire length $l$,
then macroscopic conductivity $\sigma$ saturates at values
independent on $\sigma_1$. Remarkably, for flexible wires there
are wide scaling regimes, where $\sigma \propto
(\sigma_1\sigma_2)^{1/2}$. Such dependence is known for a narrow
vicinity of percolation threshold in two dimensional isotropic
mixtures \cite{Dykhne,Efros} but, to the best of our knowledge,
has never been claimed for a broad range of parameters.

Our theory of effective conductivity of composites can be easily
applied to a completely different problem, for which the meaning
of the correlation length $\lambda$ is particularly obvious.
Namely, we speak of diffusion of proteins through the semi-dilute
system of dsDNA molecules. Many proteins have positively charged
domains on their surfaces, which provide for nonspecific
attraction to the negatively charged surface of the double helical
DNA. Such proteins stick to DNA and diffuse along DNA for some
time, then get desorbed and wander in 3D, then get adsorbed for
another tour of 1D diffusion, and so on. These phenomena are
believed to be behind the ability of proteins to locate their
specific functional targets on DNA faster than simple 3D diffusion
\cite{BWH,Halford,Klenin,first_paper}. As regards macroscopic
diffusion of proteins through semi-dilute DNA solution, we show
later in this article that this problem can be easily reduced to
that of conductivity in the composite with nanowires. As a result,
Fig. \ref{fig:rod-conductivity}, part of Figs.
\ref{fig:coil-conductivity1} and \ref{fig:coil-conductivity2}
(which is redrawn in the Fig. \ref{fig:coil-diffusion}) can all be
understood in terms of macroscopic diffusion if one uses
translation keys provided in the captions of these figures. The
results are also summarized in Table \ref{tab:table}.  Clearly, in
the case of protein diffusion $\lambda$ is the length of one tour
of protein diffusion along DNA (including episodes of activated
desorbtion if they are followed by correlated re-adsorbtion).

Our results are presented by the ``phase diagrams'' in the log-log
plane of parameters $\phi$ vs. $s=\sigma_1/\sigma_2$ shown in
Figs. \ref{fig:rod-conductivity}, \ref{fig:coil-conductivity1} and
\ref{fig:coil-conductivity2}. They specify scaling regimes of
different power laws formulae for $\sigma$ listed in the Table
\ref{tab:table}. Relatively simple phase diagram of Fig.
\ref{fig:rod-conductivity} presents results for straight wires,
while more complicated phase diagrams of Figs.
\ref{fig:coil-conductivity1} and \ref{fig:coil-conductivity2} are
constructed for semi-flexible wires characterized by a large
persistence length $p$, such that $a \ll p \ll l$.

The plan of this article is as follows. In section
\ref{sec:conductivity-stick} we first consider the relatively
simple case when each wire is straight. In this situation we
explain the main idea of our theory and identify several scaling
regimes. We then consider a more complicated case when wires are
flexible and coiled (section \ref{sec:conductivity-coil}). We
continue in section \ref{sec:diffusion} by using these results for
the macroscopic diffusion constant of proteins in semidilute DNA
system. Finally, we conclude with the discussion of other possible
applications of our work (section \ref{sec:discussion}).

In this article restrict ourselves to scaling approximation for
the conductivity and to delineating the corresponding scaling
regimes. In our scaling theory, we drop away both all numerical
factors and, moreover, also all logarithmic factors, which do
exist in the problem, because it deals with strongly elongated
cylinders.

\section{Straight wires}\label{sec:conductivity-stick}

\begin{figure}
\begin{center}
\includegraphics[width=0.45 \textwidth]{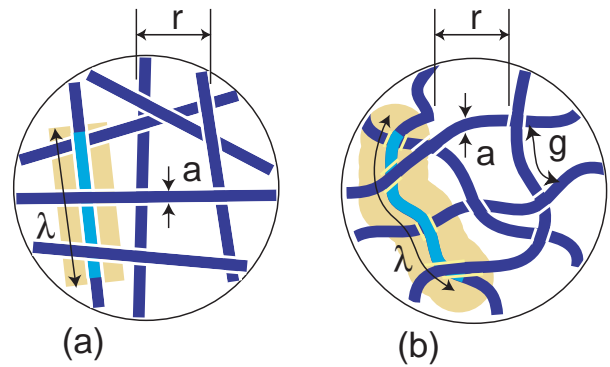}
\caption{Local view of the semidilute solutions of wires. The
correlation length along the wire is shown in lighter color than the
rest of the network. The conducting channel is shadowed. In a),
wires are straight sticks. In b), wires are flexible Gaussian coils.
If the mesh size is not longer than persistence length, so the wire
within each mesh is essentially straight. At lesser density, the
mesh size is longer than the persistence length, and then the wire
in the mesh would be wiggly. } \label{fig:network}
\end{center}
\end{figure}

In this section, we concentrate on the system of straight well
conducting sticks, such as, e.g., carbon nanotubes, suspended in the
medium of lower conductivity - see Figure \ref{fig:network}a. First
of all let us note that percolation through the wires and the direct
contacts where wires touch each other starts in such system when
volume fraction of wires exceeds a critical value which is of the
order of $\phi \sim a/l$ (see Ref. \cite{Foygel, Kolesnikov}). It is
not a coincidence that at about the same concentration it becomes
impossible to place wires randomly and isotropically \cite{Onsager}
- both percolation and nematic ordering occur at the concentration
at which there are on average about two contacts per each stick.

As stated in the introduction, we consider the range of
concentrations $a^2/l^2 < \phi < a/l$. Only in the very end we
comment on the role of percolation in our system. As long as $\phi
\ll a/l$ wires are still oriented isotropically and the contacts of
wires can still be neglected, i.e., we are below percolation
threshold; the latter means that overall macroscopic conductivity is
entirely due to the fact that the medium does conduct, $\sigma_2
\neq 0$, albeit maybe not too well. On the other hand, the spheres
containing wires already overlap strongly ($\phi > a^2/l^2$).  The
latter condition means that we deal with semidilute solution of
sticks - the system that locally looks like a network with certain
mesh size $r$ (see Figure \ref{fig:network}a).  In the scaling
sense, $r$ is the same as the characteristic radius of
density-density correlation function, and can be estimated by
noticing that one stick within one mesh makes density about $\sim
ra^2/r^3 \sim \phi$, therefore $r \sim a\phi^{-1/2}$.

\begin{figure}
\begin{center}
\includegraphics[width=0.45 \textwidth]{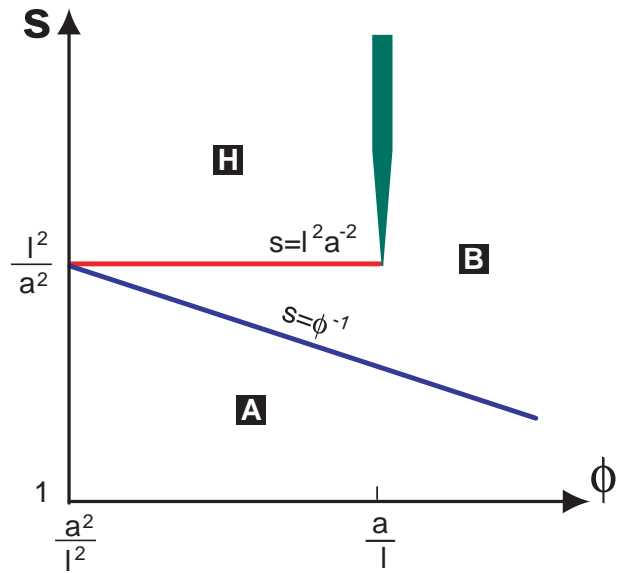}
\caption{Diagram of scaling regimes for the case of straight wires
(e.g. carbon nanotubes). Both volume fraction of the wires $\phi$
and $s=\sigma_1/\sigma_2$ axes are in the logarithmic scale. Each
line on this plane marks a cross-over between scaling regimes. The
widened line at $\phi = a/l$ shows the location of conductivity jump
around the percolation threshold. Summary of the conductivities is
provided in Table \ref{tab:table}. If $s$ is replaced by $yd$, the
diagram represents macroscopic diffusion constant of protein
discussed in section \ref{sec:diffusion}. }
\label{fig:rod-conductivity}
\end{center}
\end{figure}

Let us start with the simplest case when $\sigma_1$ is not
significantly larger than $\sigma_2$. Then current basically has
no incentive to concentrate into the wires, instead it flows all
over the place, and effective conductivity is simply that of the
medium:
\begin{equation}
\sigma \simeq \sigma_2 \ . \ \ \ \ \ {\rm (regime \ A)}
\label{eq:regimeA}
\end{equation}
In the diagram Figure \ref{fig:rod-conductivity}, this regime is
denoted as A.

It is also quite easy to find perturbative corrections to the result
Eq. (\ref{eq:regimeA}) assuming small ``conductivity contrast''
$\left(\sigma_1/\sigma_2 \right) - 1 \ll 1$.  In this case $\sigma
\simeq \sigma_2 + (\sigma_1 - \sigma_2) \phi$, which follows from
the fact that to the first order in perturbation the current lines
remain unaffected by the difference between $\sigma_1$ and
$\sigma_2$, they remain parallel straight lines (assuming for
simplicity the simplest geometry of a uniform current field), and
each of them runs through $\sigma_1$ material instead of $\sigma_2$
over the fraction $\phi$ of its length. This perturbative result
suggests that regime A continues as long as $\sigma_1 \phi \ll
\sigma_2$.

Let us now switch to the more challenging case, when $\sigma_1$ is
so much larger than $\sigma_2$ that the current is mostly carried
by the wires. To imagine the flow of current through the composite
material in this case, it is useful to think of a single current
line.  As stated in the introduction, such line typically consists
of long stretches along one wire followed by relatively short
switches from wire to wire, and the important parameter is the
typical length over which current line follows inside one wire, we
denote it as $\lambda$.

Our plan is to consider $\lambda$ as a variational parameter. That
means, we first imagine the current distribution at some given
value of $\lambda$, determine the resistance (or conductance) of
the macroscopic sample as a function of $\lambda$, and then try to
optimize $\lambda$ accordingly.  The justification of this
procedure comes from the fact that all parts of our material obey
linear Ohm's law (linear response theory), so that the requirement
of minimal dissipation, or maximal overall conductance, is
mathematically exactly equivalent to the Kirchoff's laws
determining distribution of currents in the network of resistors
\cite{Prigogine}.

To begin with, let us consider the most interesting case when
$\lambda$ is much shorter than total length of one wire, but still
larger than the mesh size $r$: $r \ll \lambda \ll l$. Consider then
a cube of the size about $\lambda$ inside our macroscopic sample. On
the one hand, overall conductivity on the scale of this cube is
already about the same as that of a macroscopic body, we denote it
$\sigma$; the resistance of one $\lambda$-size cube is, in other
words, about $1/\lambda \sigma$.  On the other hand, we can estimate
this resistance considering wires inside the cube. There are about
$\lambda^3/\lambda r^2$ of the wires crossing the cube, because the
distance between wires is about $r$ and, therefore, each wire can be
thought of as dressed in a sleeve of the thickness about $r$ and
volume about $\lambda r^2$ (see Fig. \ref{fig:network}a). The sleeve
can be thought of as a weakly leaking insulation for the wire. Each
piece of the sleeve of the length about $r$ bridges given wire to
another one through the resistance about $r/\sigma_2 r^2$, and about
$\lambda/r$ such bridges are connected in parallel, yielding overall
resistance connecting the given wire as $(1/\sigma_2r)(r/\lambda) =
1/\sigma_2 \lambda$. This is connected in series with the wire
itself, producing a conducting channel with resistance $\lambda/a^2
\sigma_1 + 1/\sigma_2 \lambda$. Since all (or sizeable fraction of
all) $\lambda^2/r^2$ conducting channels in the cube are in
parallel, we finally arrive at the cube resistance as
$(r^2/\lambda^2) \left( \lambda/a^2 \sigma_1 + 1/\sigma_2 \lambda
\right)$.  Equating this to $1/\lambda \sigma$, we arrive at the
following estimate of effective macroscopic conductivity
\begin{equation} \sigma \sim
\frac{a^2/r^2}{1/\sigma_1 + a^2/\sigma_2 \lambda^2} \ .
\label{eq:Sigma_of_lambda} \end{equation}

Let us now analyze this result.  As a function of $\lambda$,
macroscopic conductivity does not appear to have maximum at any
finite $\lambda$.  This does not necessarily mean that $\lambda$
is going to increase \textit{ad infinum}; rather, it suggests that
more accurate calculation is required to determine $\lambda$.
Luckily, such more accurate calculation is not necessary to
determine the quantity of our interest - macroscopic conductivity
$\sigma$. Indeed, as soon as $\lambda$ exceeds certain threshold,
namely
\begin{equation} \lambda > a \sqrt{\sigma_1 /
\sigma_2} \ , \label{eq:lambda_greater_than} \end{equation}
the second term in denominator of formula
(\ref{eq:Sigma_of_lambda}), which describes resistance of
wire-to-wire bridges, becomes subdominant, and must be neglected
within the accuracy of our scaling estimates.  This yields
%
\begin{equation}
\sigma \simeq \sigma_1 \phi \ . \ \ \ \ \ {\rm (regime \ B)}
\label{eq:regimeB}
\end{equation}
As expected, regimes A and B cross over smoothly on the line
$\sigma_1/\sigma_2 \sim \phi$.

In the regime B, macroscopic conductivity does not depend on
$\sigma_2$, conductivity of the medium.  This happens because
$\sigma_1/\sigma_2$ is so large that current mostly flows through
the wires, but at the same time $\sigma_1$ is not large enough to
make resistance of wires insignificant and thus unmask the
resistances of narrow gaps between wires.  This is also why the
resulting macroscopic conductivity is in this regime insensitive
to the exact value of $\lambda$: overall distance travelled by any
particular line of current through the wires scales simply as the
sample size and, to the scaling accuracy, does not depend on how
frequently current switches from wire to wire - precisely because
wires are straight.

To complete our analysis of the regime B, let us note that the
quantity $a \sqrt{\sigma_1/\sigma_2}$ which appears in the formula
(\ref{eq:lambda_greater_than}) can be understood in the following
way.  Imagine one straight wire in an infinite medium, and suppose
we somehow feed current into a certain point $0$ of this wire.
Current will flow away from $0$, mostly along the wire, but also
slightly leaking into the environment.  Due to this leaking the
current remaining in the wire will decay exponentially as we move
away from $0$, with the decay length equal to $a
\sqrt{\sigma_1/\sigma_2}$. Indeed, the length of current decay for
a single wire is estimated by the condition that the resistance of
wire over the length $\lambda$, which is about $\lambda/\sigma_1
a^2$, should be about the same as the resistance of the medium
``in perpendicular direction'', which is about $1/\sigma_2
\lambda$; equating these two returns the result
(\ref{eq:lambda_greater_than}).

Let us now switch to the next regime H, which arises because of
the finite length of sticks.  Indeed, the length $\lambda$, over
which current flows in one wire, cannot exceed the wire length
$l$. According to formula (\ref{eq:lambda_greater_than}), this
crossover happens along the border $s=l^2/a^2$.  Indeed, starting
from this $s$, we get even $a \sqrt{\sigma_1/\sigma_2} > l$, and
so $\lambda$ cannot satisfy Eq. (\ref{eq:lambda_greater_than}). To
find conductivity in the regime H, we should replace $\lambda$ by
$l$ in Eq. (\ref{eq:Sigma_of_lambda}). Since the second term in
the denominator dominates, we arrive at
\be \sigma \sim  \phi (l/a)^2 \sigma_2 \ .  \ \ \ \ \ {\rm (regime \
H)} \label{eq:regimeH} \ee
We get the same $l^2$ dependence as predicted in reference
\cite{Sarychev}. This effective conductivity has no dependence on
$\sigma_1$ because $\sigma_1$ is so high that overall resistance
is entirely concentrated in the narrow gaps where current should
switch from wire to wire. As a result, $\sigma_1$ does not enter
the formula.

If we decrease $\phi$ and look at the dilute regime $\phi <
a^2/l^2$, our scaling theory suggests that $\sigma \simeq \sigma_2$.
More accurate analysis of this regime was performed in the
work~\cite{Wang} using first order perturbation theory in powers of
$\phi$, which is applicable in the dilute regime, as long as current
field around one wire does not affect the neighboring wires. The
achievement of that work is that the authors were able to take into
account, to the first order in their perturbation theory, both
numerical coefficients and logarithmic factors proportional to $\ln
(l/a)$ (in our notations). Up to these factors, which we
systematically neglect, all our scaling results cross-over smoothly
with the results of the work~\cite{Wang}. Moreover, the simplified
expressions of perturbation results are given in the
work~\cite{Wang} in the form of three formulae, describing the
dilute solution in the ranges (in our notations) $l \gg a
\sqrt{\sigma_1/\sigma_2}$, $l \approx a \sqrt{\sigma_1/\sigma_2}$,
and $l \ll a \sqrt{\sigma_1/\sigma_2}$; this sheds an additional
light on the importance of the lengths scale $a
\sqrt{\sigma_1/\sigma_2}$ introduced by us here in the present work.

Thus, we have completed our consideration of the phase diagram up to
the concentration about $\phi = a/l$. It would be frustrating to
stop at this point, because, for example, the experiments with
suspensions of carbon nanotubes often use loadings with $\phi > a/l$
to achieve larger electrical and thermal conductivities. Let us
therefore discuss what happens to conductivity if one can manage to
create isotropic suspension with $\phi > a/l$. Suppose the length of
wire between direct contacts, or the mesh size of the percolating
network, is $\zeta$.  Then, the number of electrically parallel
wires within a cube of size $\zeta$ is of the order of
$\zeta^3\phi/(\zeta a^2)$, each with the resistance about
$\zeta/(\sigma_1a^2)$. Therefore the total resistance of the cube
scales as $1/(\sigma_1\zeta\phi)$, yielding the effective
conductivity about $\sigma_1\phi$. In the scaling sense, this is the
same $\sigma$ as that in Regime B.  Thus, regime B continues to
higher concentrations $\phi > a/l$.  Trivially, the lower bound of
this continued regime B remains to be the condition $s > 1/\phi$,
since at lower $s$ current does not concentrate in wires,
percolation gives no help.  More interestingly, percolation produces
no effect on conductivity $\sigma$ as long as $s < (l/a)^2$ because
the percolating network gives no advantage over the conducting
channels made of combination of wires and surrounding medium.
Percolation does have effect when $s > (l/a)^2$, where conductivity
as a function of concentration $\phi$ changes rapidly from $\sigma
\sim \phi (l^2/a^2) \sigma_2$ below percolation to $\sigma \sim \phi
\sigma_1$ above.  This change occurs around the threshold $\phi =
\phi_c \simeq  a/l$, over the interval of width of order $\phi_c$,
which is schematically plotted as the shaded cone in the diagram
Fig. \ref{fig:rod-conductivity}. In this range, the conductivity has
critical behavior similar to that discussed in Ref. \cite{Efros}.
The detailed structure of this range is beyond the scope of this
paper.

We should emphasize that in all our considerations we completely
disregard the resistivity of the contacts, either between a wire
and surrounding medium, or between two wires in contact.  In
particular, everything we said about percolation assumes that
whenever there is a touch between two wires, it presents an
electrical contact of vanishing resistance.  In fact, this
assumption is model sensitive and it is not necessarily good in
practical cases.  For instance, our theory predicts that above
percolation threshold, when $\phi > a/l$, effective electric
conductivity $\sigma$ grows linearly with $\phi$, independently of
$s$ being larger or smaller than $(l/a)^2$. Similar prediction
holds also for thermal conductivity. However it is not compatible
with the apparently super-linear growth of effective electrical or
thermal conductivity observed in experiments
\cite{Choi,Park,Bryning,Kilbride}. It is not clear whether
interfacial resistance can help to explain these experimental
data.

\begin{table}
\caption{The summary of macroscopic conductivities and diffusion
constants in various regimes. \label{tab:rates table}}
\begin{tabular}{|l|l|l|}

\hline Regime & $\sigma$ & $D$\\

\hline A & $\sigma_2$ & $D_3$\\

\hline B & $\sigma_1 \phi$ & $D_1 y \phi$\\

\hline C & $(p/a)\phi(\sigma_1 \sigma_2)^{1/2}$ & $(p/a)y^{1/2}\phi(D_1 D_3)^{1/2}$\\

\hline D & $(p/a)^2\phi^{3/2}(\sigma_1 \sigma_2)^{1/2}$ & $(p/a)^2y^{1/2}\phi^{3/2}(D_1 D_3)^{1/2}$\\

\hline E & $(lp/a^2)\phi \sigma_2$ & $(lp/a^2)\phi D_3$\\

\hline F & $(lp^3/a^4)\phi^2 \sigma_2$ & $(lp^3/a^4)\phi^2 D_3$\\

\hline G & $(p/a)\phi^2\sigma_1$ & no correspondence \\

\hline H & $(l^2/a^2) \phi \sigma_2$ & $(l^2/a^2)\phi D_3$\\

\hline
\end{tabular} \label{tab:table}
\end{table}

\section{Gaussian coiled wires}\label{sec:conductivity-coil}

\begin{figure}
\begin{center}
\includegraphics[width=0.45 \textwidth]{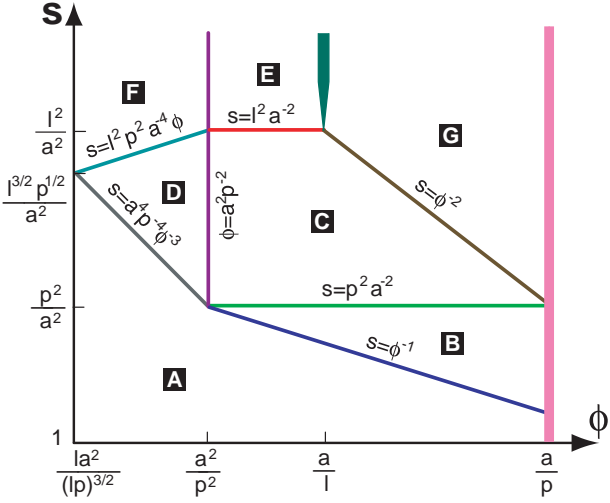}
\caption{Diagram of scaling regimes for the case of flexible
Gaussian coiled wires (e.g. conducting polymers) with length
$p<l<p^2/a$. Summary of the conductivities is provided in Table
\ref{tab:table}. } \label{fig:coil-conductivity1}
\end{center}
\end{figure}
\begin{figure}
\begin{center}
\includegraphics[width=0.45 \textwidth]{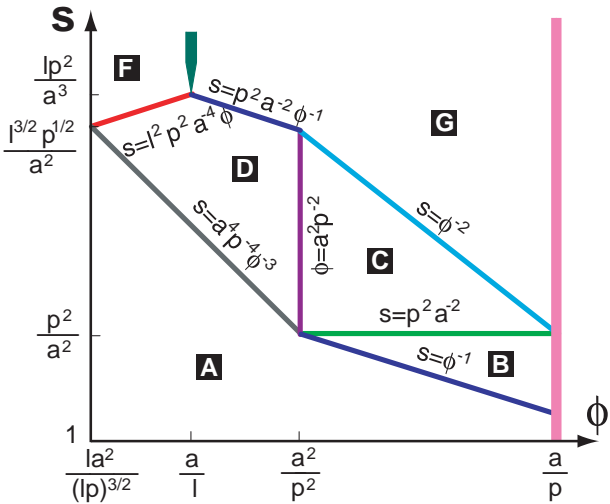}
\caption{Diagram of scaling regimes for the case of flexible
Gaussian coiled wires with length $p^2/a<l<p^3/a^2$.}
\label{fig:coil-conductivity2}
\end{center}
\end{figure}

When wires are flexible (e.g. conducting polymer), our theory
developed in section \ref{sec:conductivity-stick} needs
modifications to account for different fractal properties of the
wires. We consider the system of semi-flexible conducting polymers
as an example (see Fig. \ref{fig:network}b). We assume that each
polymer has contour length $l$, relatively large persistence length
$p$ and the radius (thickness) $a$. We assume $l \gg p \gg a$.
Throughout this work we require $l<p^3/a^2$ and this lets us
disregard the effect of excluded volume on the polymer statistics,
considering polymer coil as Gaussian. Therefore, if we take a piece
of polymer of the contour length $\lambda$, then its size in space
scales as
\be \xi \sim \left\{\begin{array}{lcr} \lambda & {\rm when} & \lambda < p \\
\sqrt{\lambda p} & {\rm when} & \lambda > p
\end{array} \right. \ . \label{eq:fractality} \ee
The overlap of coils starts when the volume fraction of polymers
exceeds volume fraction of the chain inside one coil: $\phi \sim
la^2/(lp)^{3/2}= a^2/(l^{1/2}p^{3/2})$. The nematic ordering of
wires starts at a larger volume fraction $\phi \sim a/p$. The
percolation threshold can be estimated from following argument.
Percolation happens when each wire has roughly two direct contacts
with other wires. Following Ref. \cite{Redbook}, the number $n$ of
contacts per wire is of the order of
\be n \sim \phi l/a   \label{eq:contact} \ee
(each piece of the polymer of the length $a$ has probability
$\phi$ to be in touch).  Requiring that $n$ is of the order of 1,
we obtain the percolation threshold $\phi_c \sim a/l$. Percolation
starts earlier than the nematic ordering and $\phi_c$, because
percolation requires a couple of contacts per wire, while nematic
ordering requires a contact per a smaller length of a statistical
segment ($\sim p$).  Therefore, $\phi_c$ divides into two parts
the range of concentrations when the system is semi-dilute but
isotropic $a^2/(l^{1/2}p^{3/2})< \phi< a/p$:
$a^2/(l^{1/2}p^{3/2})<\phi<\phi_c$ where we neglect the effect of
direct contacts; and $\phi_c < \phi < a/p$ where the effect of the
percolating wires must be included in our theory.

Let us start with determining the mesh size $r$ \cite{Redbook}.
Suppose that the polymer within each mesh has a contour length $g$.
It makes a density about $\sim ga^2/r^3$ which must be about the
overall average density $\phi$. Thus, $ga^2/r^3 \sim \phi$. Second
relation between $g$ and $r$ depends on whether mesh size is bigger
or smaller than persistence length $p$:
\be r \sim \left\{ \begin{array}{lcr} g & {\rm if} & g < p \\
\sqrt{gp} & {\rm if} & g >p \end{array} \right. \ . \ee
Accordingly, one obtains
\be \begin{array}{lccr} g \sim a\sqrt{\frac{1}{\phi}} \ , &
r \sim a\sqrt{\frac{1}{\phi}} & {\rm if} & \frac{a}{p} > \phi > \frac{a^2}{ p^2} \\
g \sim \frac{a^4}{\phi^{2}p^{3}} \ , & r \sim \frac{a^2}{\phi p} &
{\rm if} & \frac{la^2}{(lp)^{3/2}} < \phi < \frac{a^2}{p^2}
\end{array}  \ , \label{eq:blob_size} \ee

The upper line corresponds to such a dense network that every mesh
is shorter than persistence length and polymer is essentially
straight within each mesh (see Fig. \ref{fig:network}b). The lower
line describes much less concentrated network, in which every mesh
is represented by a little Gaussian coil. Depending on the relation
between $l$ and $p$, percolation can start before or after $\phi
\sim a^2/p^2$. Our results for these two cases are summarized in
Figs. \ref{fig:coil-conductivity1}, \ref{fig:coil-conductivity2} and
in the Table \ref{tab:table}.

When the correlation length $\lambda < p$, the polymer within
correlation length is straight. So we can directly apply what we
got for the straight wire case and obtain regimes A and B.

For other regimes with $\lambda > p$, the derivation has to be
performed from the beginning.  So, we consider a typical cube
inside the composite with size $\xi$ such that every polymer
enclosed in this cube has contour length of the order of
correlation length $\lambda$. There are about
$\frac{\xi^3}{r^3(\lambda/g)}$ electrically parallel conducting
channels in this cube (because $\lambda/g$ is the number of meshes
visited by one wire, and $r^3$ is the volume of each such mesh).
Each channel consists of the wire itself, with resistance
$\lambda/(\sigma_1a^2)$, and the wire is connected in series with
the group of $\lambda/g$ parallel connected bridges, each of
resistance $r/(\sigma_2r^2)$. Thus, the resistance of the cube
scales as
\be \frac{\lambda/(\sigma_1 a^2) +g/(\sigma_2 \lambda r)}{g
\xi^3/(\lambda r^3)} \ , \ee
which should be equated to $1/(\sigma\xi)$. Therefore we obtain:
\be \sigma \simeq  \frac{a^2 g/r^3}{\lambda^2/(\sigma_1 \xi^2) +
ga^2/(\sigma_2 r \xi^2)} \ . \label{eq:sigma_of_lambda_fractal}
\ee
Once again, if the wire remains straight over the length
$\lambda$, so that $\xi \simeq \lambda$, we are back at the
situation described by formula (\ref{eq:Sigma_of_lambda}), and we
can reproduce the corresponding result for the regime B,
(\ref{eq:regimeB}).  More interestingly, we can now consider the
case when $\lambda > p$ and the polymer of the length $\lambda$ is
the Gaussian coil: $\xi \sim \sqrt{\lambda p}$.  In this case we
obtain
\be \sigma \sim \frac{pa^2g/r^3}{\lambda/\sigma_1+a^2g/(\sigma_2 r
\lambda)} \ . \label{eq:conductivity-coil} \ee
Now conductivity has the well defined maximum at the well defined
value of $\lambda$:
\be \lambda \sim a \left( \frac{\sigma_1 g}{\sigma_2 r}
\right)^{1/2} \ . \label{eq:lambdacoil} \ee
Not coincidentally, this result for the correlation length
$\lambda$ corresponds to equating two terms in the denominator of
Eq. (\ref{eq:conductivity-coil}) - the resistance of wire with
correlation length $\lambda$ and the resistance of the surrounding
``sleeve'' of thickness $r$ in the medium.

Plugging $\lambda$ from Eq. (\ref{eq:lambdacoil}) back into Eq.
(\ref{eq:conductivity-coil}) and applying Eqs.
(\ref{eq:blob_size}), we obtain the following two scaling regimes:

Regime C, where polymer on the scale $\lambda$ is Gaussian (lower
line in the Eq. (\ref{eq:fractality}), but the polymer within each
mesh is still straight;

Regime D, where polymer is Gaussian even within each mesh (lower
lines in the Eqs. (\ref{eq:blob_size})).

When $\lambda$ reaches the entire length of the polymer $l$, we
should replace $\lambda$ by $l$ in Eq.
(\ref{eq:conductivity-coil}). Then the second term in the
denominator dominates. Since we have two different kinds of meshes
represented by formulae (\ref{eq:blob_size}), we have two more
regimes:

Regime E, where $\lambda = l$ and the polymer within each mesh is
straight (upper lines in the Eq. (\ref{eq:blob_size}));

Regime F, where $\lambda=l$, but the polymer within each mesh is
Gaussian (lower lines in the Eq. (\ref{eq:blob_size})).

We should emphasize that regime E exists only in the case where $l <
p^2/a$ (see Fig. \ref{fig:coil-conductivity1}). If $l>p^2/a$,
percolation starts so early that the polymer within each mesh is a
Gaussian coil. This case is plotted in Fig.
\ref{fig:coil-conductivity2}.

When we increase $\sigma_1/\sigma_2$, the effective conductivity
grows from $\sigma_2$ (regime A) and finally saturates in regimes
E and F with values having no dependence on $\sigma_1$. As we
discussed, for regime A, transport through wires is not at play
while in regimes E and F, $\sigma_1$ is so large compared to
$\sigma_2$ that the wires are effectively
\emph{"super-conducting"}. More interestingly, for broad ranges of
$\phi$ and $\sigma_1/\sigma_2$ (regimes C and D), $\sigma$ is
proportional to $\sqrt{\sigma_1 \sigma_2}$. Such dependence is
known for a narrow vicinity of percolation threshold in isotropic
mixtures \cite{Dykhne} but, as far as we know, it has never been
noticed for a broad range of parameters. The width of the range
grows as $l^2$.

When the volume fraction is larger than the percolation threshold,
the effect of percolating wires can not be neglected. If
$\sigma_1/\sigma_2$ is relatively small, the transport of current is
mainly realized through the conducting channel we have discussed.
But when $\sigma_1/\sigma_2$ is large enough, percolation through
the directly connected wires dominates. The crossover is determined
by equating the conductivity due to the untouched wires and
surrounding medium and the conductivity due to percolating wires. We
have already calculated the first conductivity. The later one can be
calculated by the following argument. Let us denote the length of
wire between contacts by $\zeta$. It can be estimated as $\zeta \sim
l/(\phi/\phi_c) \sim a/\phi$. Because we require $\phi < a/p$,
$\zeta$ is larger than the persistence length $p$ and thus the
distance it covers in space scales as $\sim (\zeta p)^{1/2}$. Within
a cube with size $(\zeta p)^{1/2}$, there are $(\zeta
p)^{3/2}\phi/(\zeta a^2) \sim (p/a)^{3/2}\phi^{1/2}$ electrically
parallel wires. So the conductance scales as $\sigma_1pa/\zeta$. It
can be also expressed using effective conductivity, it is $\sigma
(\zeta p)^{1/2}$. Comparing these two conductances, we obtain
$\sigma \sim (p/a)\phi^2\sigma_1$, which is the effective
conductivity in regime G. It crosses over smoothly to the regimes E
and F (E only exists for case $l<p^2$, which is represented in Fig.
\ref{fig:coil-conductivity1}). One can also obtain the border by
equating the correlation length $\lambda$ to the length between
direct contacts $\zeta$. Since we assume the current can switch
wires freely at the contacts, $\lambda$ can not grow above $\zeta$.
On the other hand in the scaling approach we use, there is a
conductivity jump around the percolation threshold $\phi_c \sim
a/l$, which is plotted as the widened line in both figures. Actually
the jump of conductivity is eliminated when we consider the critical
behavior of conductivity at $\phi-\phi_c \ll \phi_c$ but our scaling
theory is not designed to see such details.

\section{Macroscopic diffusion constant of proteins in semidilute DNA
system} \label{sec:diffusion}

\begin{figure}
\begin{center}
\includegraphics[width=0.45 \textwidth]{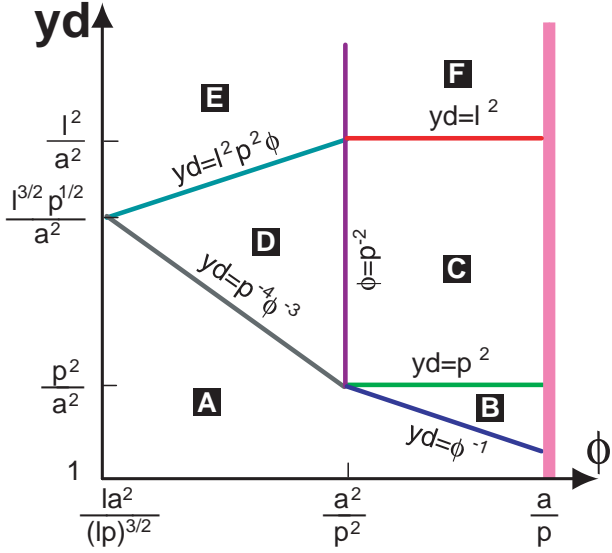}
\caption{Diagram of scaling regimes for the case of protein
diffusion in Gaussian coiled DNA solution. Summary of the
macroscopic diffusion constants is provided in Table
\ref{tab:table}. } \label{fig:coil-diffusion}
\end{center}
\end{figure}

The theory we have developed for the effective conductivity of
composites can be used to study the macroscopic diffusion constant
of the nonspecific DNA-binding proteins in semidilute DNA
solutions.

For simplicity, we make the following assumptions: (i) protein can
be non-specifically adsorbed on any place of DNA; (ii)
non-specific adsorption energy $\epsilon$, or the corresponding
constant $y = e^{\epsilon/k_BT}$, are the same everywhere along
DNA (sequence-independent); (iii) every protein molecule has just
one surface patch capable to stick to the DNA, so proteins do not
serve as cross-linkers for the DNA; (iv) non-specifically bound
protein can diffuse along DNA with the diffusion coefficient
$D_1$, while protein dissolved in surrounding water diffuses in 3D
with diffusion constant $D_3$; (v) while protein is diffusing, the
DNA remains immobile.

%
%
%
%

To make the dictionary of translation between conductivity and
diffusion languages, the easiest way is to step up the generality
in writing down the expressions for current density ${\mathbf j}$
in either conductivity or diffusion problem. In both cases, as
long as we consider stationary process, current is subject to the
no-divergence condition: ${\rm div} {\mathbf j} = 0$.  For the
electric current driven by the potential gradient, Ohm's law reads
${\mathbf j} = - \sigma ({\mathbf x}) \nabla \phi$; and for the
the diffusion problem, current driven by the gradient of total
chemical potential is described by similar Smoluchowski equation:
${\mathbf j} = - D({\mathbf x}) c({\mathbf x}) \nabla \left( \ln
c({\mathbf x}) - \epsilon({\mathbf x})/k_BT \right)$. Here, we
assume for a moment, that electrical conductivity, diffusion
coefficient, protein concentration $c$, and binding energy
$\epsilon$ or $y=e^{\epsilon/k_BT}$ have all some general
dependence of space coordinates ${\mathbf x}$. In fact, for our
case, this space dependence is very simple: within narrow regions
along the wires or along DNA, we have $\sigma({\mathbf x}) =
\sigma_1$, and similarly $D({\mathbf x})=D_1$ and $y({\mathbf x})
=y$ (remember that $y = e^{\epsilon/k_BT}$); for all other places
${\mathbf x}$ we have $\sigma({\mathbf x}) = \sigma_2$,
$D({\mathbf x})=D_3$ and $y({\mathbf x}) =1$.  As regards
concentration, diffusion equation also implies (since chemical
potential is continuous) that locally there is the equilibrium
relation between the 1D concentration of non-specifically adsorbed
proteins, $c_1$, and concentration of proteins remaining free in
the nearby solution $c_3$:
\be c_1/(c_3a^2) = y \ , \label{eq:concentration} \ee
where $a$ is the length scale such that $c_1/a^2$ is the 3D
concentration of proteins within the region around DNA where
proteins are adsorbed. 
Comparing the equations, we see that complete mapping is achieved
by the substitutions $\sigma_1 \leftrightarrow D_1 c_1/a^2$,
$\sigma_2 \leftrightarrow D_3c_3$. Similarly writing the effective
macroscopic equations in terms of macroscopic conductivity
$\sigma$ and macroscopic diffusion coefficient $D$, one finds
$\sigma \leftrightarrow D c_3$.  In terms of more convenient
dimensionless quantities, and taking into account the local
adsorbtion equilibrium (\ref{eq:concentration}), these rules read:
\be \frac{\sigma_1}{\sigma_2} \leftrightarrow \frac{D_1}{D_3}y \ \
, \ \ \frac{\sigma}{\sigma_2} \leftrightarrow \frac{D}{D_3} \ .
\label{eq:ratio_of_sigma} \ee
We can, therefore, directly address macroscopic diffusion based on
our results for macroscopic conductivity. Substituting Eq.
(\ref{eq:ratio_of_sigma})
into our results for $\sigma$, we obtain the macroscopic diffusion
constants of the proteins in the DNA solution expressed in terms of
$D_1$ and $D_3$ for all the regimes except regime G, for which the
applicability of percolation results to the DNA case is doubtful.
\footnote{As we mentioned, even for the wires the idea of direct
contact is very much model-dependent. It is even more so for the
protein and DNA case, because in this case ``contact between wires''
should mean the possibility for the protein to switch from one DNA
to the other without activation. It might be possible in some
systems but impossible in others; besides, there are quite a few
other effects which are beyond our theory, such as excluded volume
constraints for the proteins which becomes significant when two DNA
pieces are close by - protein may have difficulties diffusing along
one of them, like a big truck under a low bridge on a highway.  We
do not consider all these questions in this paper, and only consider
DNA system well below percolation threshold.}


Thus, Fig. \ref{fig:rod-conductivity} can be used for the
macroscopic diffusion constants for the straight DNA case if we
replace $\sigma_1/\sigma_2$ by $yD_1/D_3$. However, for the
gaussian coiled DNA case, without regime G, Figs.
\ref{fig:coil-conductivity1} and \ref{fig:coil-conductivity2}
should be modified. Resulting phase diagram is shown in Fig.
\ref{fig:coil-diffusion}. One can easily get this figure from Fig.
\ref{fig:coil-conductivity1} by removing the borders the regime G
makes with other regimes and extending the border line between
regimes E and C to the right boundary of the phase diagram. The
results for various scaling regimes are also summarized in the
right most column of Table \ref{tab:table}.

Measuring macroscopic diffusion of proteins is a promising way to
test our predictions.  It is therefore useful to comment a little
deeper on the nature of macroscopic diffusion coefficient $D$. The
way it is defined above is adequate for a macroscopic experiment,
because $D$ establishes the proportionality between the flow of
proteins and the gradient of concentration of \emph{dissolved}
proteins.  In such experiment, the presence of a large number of
proteins adsorbed on DNA is not directly relevant. However, in a
different experiment, for instance, in tracking the random walks of
a single protein molecule, a different diffusion coefficient,
$\tilde{D}$, will be relevant, such that $\tilde{D} c = D c_3$,
where $c$ is total concentration of proteins, including the adsorbed
ones: $c= (c_1/a^2) \phi + c_3 (1-\phi)$.  Using
(\ref{eq:concentration}), one then gets $\tilde{D} = D / \left( 1 -
\phi + y \phi \right)$. The difference between these two diffusion
coefficients is marginal when absorbtion is weak ($y \ll 1$ and
$\phi \ll 1$), but it becomes very much pronounced when the
absorbtion is strong, or $y$ is large: $\tilde{D} \simeq D/ (y \phi)
\ll D$. This result has simple physical meaning: every particular
protein will be adsorbed most of the time, so its diffusive motion
will be slow, but overall flow of proteins will not be that slow
because of a large number of proteins.

Using the macroscopic diffusion constant, we can also re-derive the
results of the work \cite{first_paper} concerning the rates of
protein searching for specific places on globular DNA. Thus,
measuring $D$ or $\tilde{D}$ is another way to verify the results of
Ref. \cite{first_paper} in those crowded regimes.

\section{Conclusion}\label{sec:discussion}

In this paper we studied a plethora of different scaling regimes
for conductivity of a suspension of wires in poorly conducting
medium. Our results are applicable to suspensions of metallic
wires in poorly conducting medium at room temperature. In this
case our generic description of the system only by two macroscopic
local conductivities is valid, because typically the surfaces of
nanowires are so dirty that any surface barrier for electrons of
the metal is sufficiently well conducting due to hopping through
localized states.

We also mention carbon nanotube suspensions as a possible
application of our theory. In this case one may worry about the role
of the surface resistance on the nanotube-medium interface, so that
our theory strictly speaking only estimates the effective
conductivity from above. Including surface resistance or allowing
for influence of environment on conductivity of nanotubes would
require new parameters in the theory, making it much more
complicated. We believe that both real and computer experiments
should be first compared with the simplest and generic model
presented here in this paper before one starts developing more
complicated theories.

The serious advantage of our generic theory is that it can be
applied in a variety of different problems beyond electric
conductivity, for instance, to thermal conductivity of well
thermally conducting wires in a weaker thermally conducting
medium, to macroscopic dielectric constant of suspended metallic
wires and to wires with large magnetic susceptibility. We also
applied our theory to the calculation of the macroscopic diffusion
constant of the nonspecific DNA-binding proteins in semi-dilute
DNA solution.

The latter application is also promising in terms of computational
tests of our theory along the lines of the recent work
\cite{Klenin,Klenin_condmat}.

\acknowledgements We are grateful to D. J. Bergman, M. Foygel, M.
Fogler, V. Noireaux and A. K. Sarychev for useful discussions. The
work of AG was supported in part by the MRSEC Program of the
National Science Foundation under Award Number DMR-0212302. The work
of AG and TH was supported in part by the grant from the US-Israel
Binational Science Foundation, Award 2004251.

\end{document}